\documentclass[12pt]{iopart}
\usepackage{iopams}

\usepackage{graphicx}
\usepackage{amsfonts}
\usepackage{float}
\usepackage{epsfig}
\usepackage{graphicx}
\usepackage{dcolumn}
\usepackage{bm}
\usepackage{times}
\usepackage{xcolor}

\begin{document}

\title[]{Optimal resource diffusion for suppressing disease spreading in
multiplex networks}

\author{Xiaolong Chen$^{1,2}$, Wei Wang$^{2,3}$, Shimin Cai$^{1,2}$,
H. Eugene Stanley$^{5}$, Lidia A. Braunstein$^{6,5}$}

\address{$^{1}$ Web Sciences Center, University of Electronic
Science and Technology of China, Chengdu 611731, China}

\address{$^{2}$ Big data research center, University of Electronic
Science and Technology of China, Chengdu 610054, China}

\address{$^{3}$ College of Computer Science and Technology,
Chongqing University of Posts and
Telecommunications, Chongqing 400065, China}

\address{$^{5}$ Center for Polymer Studies and Department of Physics,
  Boston University, Boston, Massachusetts 02215, USA}

\address{$^{6}$ Instituto de Investigaciones F\'{i}sicas de Mar del Plata
  (IFIMAR)-Departamento de F\'{i}sica, Facultad de Ciencias Exactas y
  Naturales, Universidad Nacional de Mar del Plata-CONICET, Funes 3350,
  (7600) Mar del Plata, Argentina.}

\ead{wwzqbx@hotmail.com}
\vspace{10pt}

\begin{abstract}
\noindent
Resource diffusion is an ubiquitous phenomenon, but how it impacts
epidemic spreading has received little study. We propose a model that
couples epidemic spreading and resource diffusion in multiplex
networks. The spread of disease in a physical contact layer and the
recovery of the infected nodes are both strongly dependent upon
resources supplied by their counterparts in the social layer. The
generation and diffusion of resources in the social layer are in turn
strongly dependent upon the state of the nodes in the physical contact
layer. Resources diffuse preferentially or randomly in this model.
To quantify the degree of preferential diffusion,
a bias parameter that controls the resource diffusion is proposed.
We conduct extensive
simulations and find that the preferential resource diffusion can change
phase transition type of the fraction of infected
nodes. When the degree of interlayer correlation is below a critical
value, increasing the bias parameter changes the phase transition from
double continuous to single continuous. When the degree of interlayer
correlation is above a critical value, the phase transition changes from
multiple continuous to first discontinuous and then to hybrid. We find
hysteresis loops in the phase transition. We also find that there is an
optimal resource strategy at each fixed degree of interlayer correlation
where the threshold reaches a maximum and under which the disease can be
maximally suppressed. In addition, the optimal controlling parameter
increases as the degree of inter-layer correlation increases.
\end{abstract}

%
\section{Introduction}\label{sec:intro}

\noindent
Epidemic spreading is an important topic in complex-systems theory
\cite{pastor2015epidemic} and much research on its underlying dynamics
has been conducted in recent years. Although a strong focus has been on
the theoretical analysis of epidemic spreading \cite{pastor2001epidemicprl,gomez2010discrete},
research has also included the control and
prediction of disease outbreaks
\cite{givan2011predicting,wang2016unification},
the spread of rumors \cite{moreno2004dynamics,trpevski2010model}, and
the propagation of computer viruses
\cite{newman2002email,balthrop2004technological}.
As more and more infectious diseases such as Severe acute respiratory syndrome
(SARS) \cite{lau2005severe}, Ebola virus \cite{team2015west} have
brought disasters to humans, how to constrain the global pandemics
has been one of most important and pressing challenges. In recent years,
many immunization strategies have been proposed for containing and limiting
epidemics. Traditional immunization strategies fall into two
categories. The first category includes topology-based strategies, such
as random immunization \cite{pastor2002immunization,zuzek2015}, targeted
immunization
\cite{dezsHo2002halting,holme2002attack,buono2015immunization},
acquaintance immunization \cite{cohen2003efficient}, and graph
partitioning \cite{chen2008finding}. Recent successes have used a
targeted destruction of the potential transmission network before an
outbreak occurs. ``Super-blockers'' are
identified and immunized to efficiently
break network connectivity \cite{clusella2016immunization}.
The second category includes those that focus on the dynamics of the
diffusion of information about the disease, such as information-driven
vaccination patterns
\cite{funk2009spread,ruan2012epidemic,granell2013dynamical}. Another
research topic in epidemic spreading is developing optimal strategies
of deploying limited resources such that the epidemic outbreak can be most
efficiently suppressed \cite{tragler2001optimal,lokhov2017optimal,chen2017optimal}.

Most research on immunization strategies and optimal resource deployment
assumes that available resources are fixed, static, and exist
independent of the dynamic epidemic process, but in real-world scenarios
the amount of such available resources as drugs, medical personnel, and
financial support are strongly affected by the evolution of the
disease. For example, a pandemic, e.g., the Ebola virus disease (EVD)
\cite{team2015west}, can quickly become an enormous economic burden to a
region \cite{gallup2001economic}, and even after the disease has been
brought under control the economic recovery of the region is slow
\cite{berke1993recovery}. Much recent research has examined how dynamic
changes in resources affect the dynamics of epidemic spreading.
Some research has focused on public resources
\cite{bottcher2015disease,bottcher2016connectivity,chen2016critical}.
For example, Ref.~\cite{bottcher2015disease} describes how resource
constraints caused by the outbreak of disease affect the dynamics of the
epidemic. They assume that healthy individuals in the system provide the
needed resources, and that the number of these healthy individuals
decreases as the infection rate increases.
Reference~\cite{chen2016critical} finds that there is a critical amount
of invested public resource needed to constrain the spread of a disease,
and when that amount is larger than the critical value, the disease can
be suppressed. If it is not, the fraction of infected individuals can
quickly increase. Other
researchers assume that real-world infected individuals cannot
always receive public resources and must seek help from friends in their
social circles \cite{chen2017suppressing}, and that understanding this
phenomenon is important in controlling an
epidemic. Reference~\cite{chen2017suppressing} examines how social
supports affect epidemic spreading in a double-layer multiplex network
in which one layer is the pattern of resource allocation and the other
is of epidemic spreading. They find a hybrid transition in the fraction
of infected nodes that exhibits properties of both continuous and
discontinuous phase transitions.

Although the above literature examines the dynamic evolution of
resources and their influence on epidemic spreading, it overlooks the
phenomenon of resource diffusion among individuals. Such resources as
economic wealth constantly flow among individuals. An important topic
for research involves the so-called ``Matthew effect''
\cite{perc2014matthew} in which the flow of economic wealth tends to
make the rich richer. This is relevant because infected individuals with
wealth tend to receive better treatment and have a higher probability of
recovering than those without.

To investigate the properties of resource diffusion and how it impacts
disease spreading, we examine its multiplex structure
\cite{mucha2010community,de2016physics,chen2017suppressing}. We form a
two-layer multiplex network of $N$ nodes. Each node in one layer has a
counterpart in the other layer. The
structure of the two layers can differ. For example, a person may have
one group of friends with whom they have regular face-to-face contact
and another group of friends in the on-line world
\cite{szell2010multirelational}.

Here we investigate how resource diffusion affects the dynamics of
epidemic spreading in two-layer multiplex networks. We assume that
resources diffuse among nodes in the social layer $\mathcal{S}$, and
that the disease spreads in the physical contact layer
$\mathcal{C}$. Because the diffusion of resources among nodes in layer
$\mathcal{S}$ can be either preferential or random, we introduce a bias
parameter $\alpha$ that controls the diffusion. When the nodes are
healthy they can generate new resources. The recovery of infected nodes
in layer $\mathcal{C}$ depends on the resources of their counterparts in
layer $\mathcal{S}$. Through simulations we find that the preferential
diffusion of resources can change the phase transition type of the
fraction of infected nodes at the steady state $\rho(\infty)$. When the
degree of interlayer correlation $r$ is below a critical value $r_c$,
and the initial fraction of infected nodes $\rho(0)$ is large, i.e.,
$\rho(0)=0.99$, the phase transition $\rho(\infty)$ changes from two
continuous phase transitions to a single continuous transition as
$\alpha$ increases. In addition, there are two hysteresis loops
accompanying the two phase transitions when $\alpha$ is below a critical
value $\alpha_c$, and one hysteresis loop when $\alpha>\alpha_c$.  When
$r>r_c$, the phase transition of $\rho(\infty)$ changes from multiple
(when $\alpha$ is too large or too small) to discontinuous, and then to
hybrid, with a initial continuous transition followed by a discontinuous
transition.  There is always a single hysteresis loop. Note that there
is an optimal strategy of resource diffusion under which the disease can
be most effectively suppressed, and the threshold reaches a maximum.

\section{MODEL} \label{sec:model}

\subsection{The social-contact double layer network}
We model the
coupling of the dynamics of disease spreading and resource diffusion in
a double-layer multiplex network. Each individual has links with
colleagues or coworkers in the physical contact layer and also with
friends in the social relation layer. We construct the double-layer
multiplex network model using the uncorrelated configuration model to
independently generate layers $\mathcal{S}$ and $\mathcal{C}$
\cite{catanzaro2005generation}. These two subnetworks have the same
number of nodes $N$, and there is a one-to-one correspondence between
nodes in the two layers. Each layer also has its own internal
structure. In an uncorrelated double-layer network, the node degrees in
the first layer are independent of the nodes degrees in the second. Thus
a high-degree node in the first layer does not not necessarily have a
corresponding high-degree node in the second.  In contrast, in a
correlated double-layer network the node degrees in one layer are
somewhat dependent on the node degrees in the other layer.
Quantitatively, we use the Spearman rank correlation coefficient $r$
\cite{lee2012correlated,wang2014asymmetrically} in which $r\in[-1,1]$ to
characterize the degree correlation between the two layers. For example,
when $r>0$ the two layers are positively correlated. A larger $r$ value
indicates a higher probability that a high-degree node in the first
layer matches a high-degree node in the second layer. In contrast, when
$r<0$ the two layers are negatively correlated.  A smaller value of $r$
indicates a higher probability that a high-degree node in the first
layer matches a low-degree node in the second layer.  The topological
structure of the two layers are encoded in the two adjacency matrices
$A^{\mathcal{S}}=\{a_{ij}^{\mathcal{S}}\}$ and
$A^{\mathcal{C}}=\{a_{ij}^{\mathcal{C}}\}$, respectively.  If nodes $i$
and $j$ are connected by a link in layer $\mathcal{S}$ ($\mathcal{C}$),
$a_{ij}^{\mathcal{S}}=1$ ($a_{ij}^{\mathcal{C}}=1$), otherwise
$a_{ij}^{\mathcal{S}}=0$ ($a_{ij}^{\mathcal{C}}=0$).

\subsection{Coupling disease spreading and resource diffusion}
To examine how resource diffusion affects epidemic spreading we propose a
resource-based susceptible-infected-susceptible (rSIS) model to describe the
epidemic spreading in layer $C$. In the rSIS model, each node can be
either susceptible or infected. The recovery process of the infected
nodes depends on the resources of their counterparts in layer
$\mathcal{S}$. We denote ${\rho_i}(t)$ to be the the probability that
node $i$ is infected at time $t$, and $\rho(t)$ the fraction of infected
nodes at $t$, which is determined by averaging over the infection
probability of all nodes
\begin{equation}
  \rho(t)=\frac{1}{N}\sum_{i=1}^{N}\rho_i(t).
  \label{infsizet}
\end{equation}
Here $\rho(\infty)$ is the fraction of infected nodes when
$t\rightarrow\infty$.

We first randomly select a fraction of $\rho(0)$ nodes to be seeds
(infected nodes) and leave the remaining nodes in the susceptible state.
At each time step the infected nodes transmit the disease to susceptible
neighbors at an infection rate $\beta$. The recovery of infected nodes
is dependent upon resources supplied by their counterparts in layer
$\mathcal{S}$.

Because resources can promote the recovery of infected nodes, we
consider that when a node in layer $\mathcal{S}$ has greater resources
the corresponding node in layer $\mathcal{C}$ will have a higher
recovery rate.  We denote $\mu_i(t)$ the recovery rate of node $i$ at
time $t$, which is a monotonically increasing function of the resource
quantity owned by the counterpart of $i$ in layer $\mathcal{S}$.  Note
that $\mu_i(t)$ is a constant value for all nodes in the classical SIS
model.  Specifically, $\mu_i(t)$ can be expressed
\begin{equation}
 \mu_i(t)= 1-(1-\mu_0)^{\omega_i(t)},
\label{rateRecovery}
\end{equation}
where $\mu_0$ is the basic recovery rate, which we here fix at
$\mu_0=0.1$, and $\omega_i(t)$ is the accumulated resources of the
counterpart of node $i$ in layer $\mathcal{S}$ at time $t$.

The resource diffusion in layer $\mathcal{S}$ is dependent upon the
state of nodes in layer $\mathcal{C}$. At each time step, if node $i$ in
layer $\mathcal{C}$ remains in the S state, the corresponding node in
layer $\mathcal{S}$ generates a new unit of resource. At the same time,
depending on the sign of $\alpha$, it preferentially transfers one unit
of resource to one of its neighbors (the target neighbor). Note that the
target neighbor is chosen independent of its state, but the target node
does not transmit resources to neighbors if it is not in the S state.

We denote $\phi_{i\rightarrow j}$ the resource transfer probability
from node $i$ to $j$ and assume that this transfer probability is related to the
degree of $j$. Then $\phi_{i\rightarrow j}$ is
\begin{equation}
  \phi_{i\rightarrow
    j}=\frac{(a_{ij}^{\mathcal{S}}+\delta_{ij})k_j^{\alpha}}{\sum_{\ell}a_{\ell
      i}^{\mathcal{S}}k_{\ell}^{\alpha}+k_i^{\alpha}},
\label{transProb}
\end{equation}
where $\delta_{ij}=1$ if $i=j$, otherwise $\delta_{ij}=0$. The
parameter $\alpha$ allows us to tune the degree of preference. When
$\alpha>0$, $\phi_{i\rightarrow j}$ is positively related to the
degree of $j$ and a high-degree neighbor has a high probability of
being selected, but when $\alpha=0$, every neighbor of node $i$ has
the same probability of being selected. Note that when $i=j$ node $i$
retains the unit of resource during the current time step.  The
resources $\sigma_j(t)$ that node $j$ acquires from healthy neighbors
at time $t$, can be written
\begin{equation}
  \sigma_j(t)=\sum_{i=1}^{N}a^{\mathcal{S}}_{ij}\phi_{i\rightarrow
    j}(1-\rho_i(t)).
  \label{expectRes}
\end{equation}
When node $i$ in layer $\mathcal{C}$ is in the I state, the
corresponding node in layer $\mathcal{S}$ does not generate a new
resource unit nor does it transfer a resource unit to its neighbors. The
accumulated resources of the counterpart of node $i$ in layer $S$ are
consumed. For simplicity, we assume that infected nodes consume the all
resources of their counterparts. Thus $\omega_i(t)$ returns to $0$ at
the current time step. The susceptible nodes store the resources to
distribute to neighbors or recover when they are infected in the
following time.

We use synchronous updating \cite{pastor2001epidemicprl} to simulate the
coupled dynamic process of disease spreading and resource diffusion. At
each time step with a probability $\beta\Delta t$ a susceptible node is
infected by one of its infected neighbors. Simultaneously, infected
nodes recover with a probability $\mu_i(t)\Delta t$, where $i=1...N$. We
set a time step $\Delta t=1$ and run each simulation sufficiently long
to ensure that the system enters a steady state in which either no nodes
are infected or the number of infected nodes fluctuates within a small
range.

\section{Simulation results for uncorrelated networks}
Here we examine how preferential resource diffusion affects disease
spreading in uncorrelated double-layer networks. We focus on networks
with a heterogeneous degree distribution because many networked
systems in both nature and technological applications are complex and
have a heterogeneous degree distribution
\cite{viswanath2009evolution,adamic2000power}. We use an uncorrelated
configurational
model~\cite{molloy1995critical,catanzaro2005generation} to build a
double-layer network in which the degree distribution is $P(k)\sim
k^{-\gamma_{\mathcal{S}}}$ for layer $\mathcal{S}$ and is $P(k)\sim
k^{-\gamma_{\mathcal{C}}}$ for layer $\mathcal{C}$, where
$\gamma_{\mathcal{S}}$ and $\gamma_{\mathcal{C}}$ are the power
exponents.
We fix both values of the power exponential at
$\gamma_{\mathcal{S}}=\gamma_{\mathcal{C}}=2.2$, and both
$\gamma_{\mathcal{S}}$ and $\gamma_{\mathcal{C}}$ are denoted to
$\gamma$ if there is no other special statement. To avoid degree
correlations between two layers, each layer is made
independent. Because the simulations are time consuming, we set
the system size to $N=N_{\mathcal{S}}=N_{\mathcal{C}}=5000$. For the
maximum degree we use the structural cut-off $k_{\rm max}\sim\sqrt N$
\cite{boguna2004cut} and set the minimum degree at $k_{\rm min} = 2$
\cite{cohen2003structural}. To determine the epidemic threshold, we
use a susceptibility measurement
\cite{binder1993monte,ferreira2012epidemic}
\begin{equation}
  \chi=N\frac{\langle\rho(\infty)^{2}\rangle-\langle
\rho(\infty)\rangle^{2}}{\langle\rho(\infty)\rangle},
  \label{chi}
\end{equation}
where $\langle\cdots\rangle$ is the ensemble averaging, and $\chi$
exhibits peaks at the transition points if they exist.

\begin{figure}[t]
\begin{center}
\includegraphics[width=1.0\linewidth]{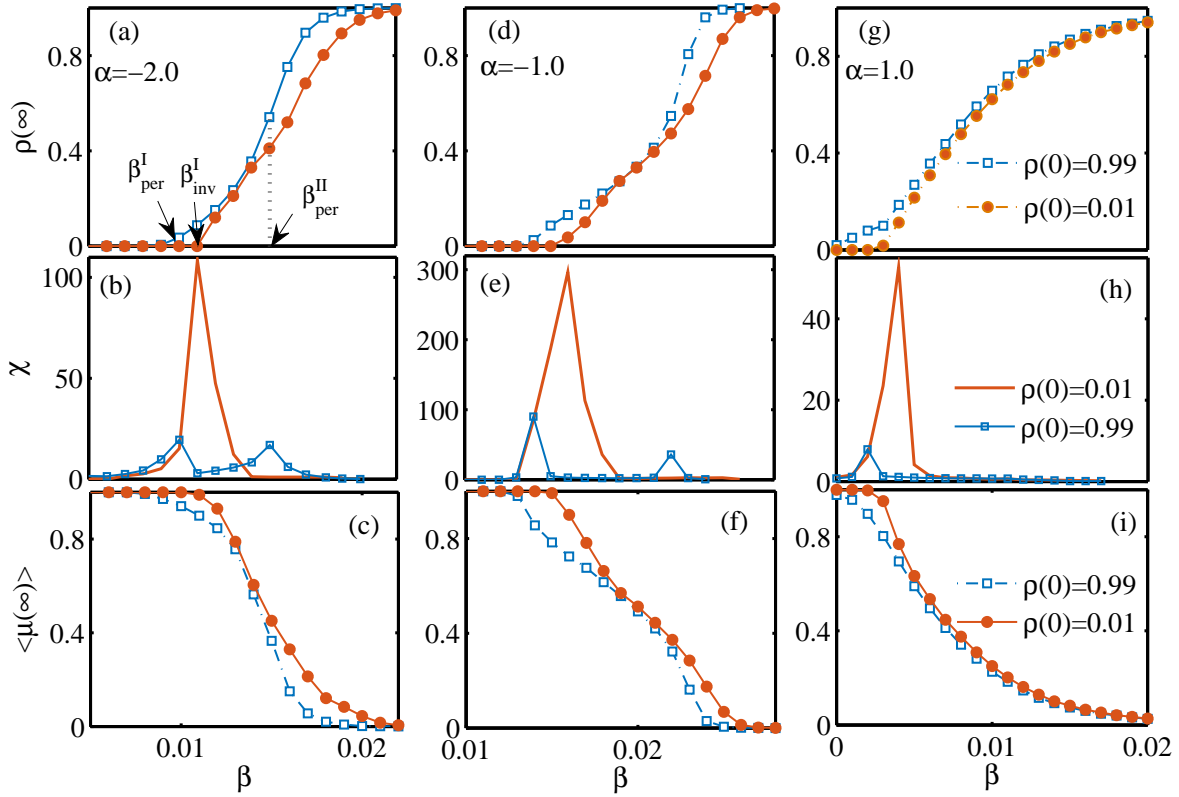}
\caption{(Color online) Influence of preferential resource diffusion on
  disease spreading. Fraction of infected nodes $\rho(\infty)$ as a
  function of $\beta$ for $\alpha=-2.0$ (a), $\alpha=-1.0$ (d) and
  $\alpha=1.0$ (g) respectively. Susceptibility $\chi$ as a function of
  $\beta$ for $\alpha=-2.0$ (b), $\alpha=-1.0$ (e) and $\alpha=1.0$
  (h). Average recovery rate at the steady state
  $\langle\mu(\infty)\rangle$ as a function of $\beta$ for the
  corresponding $\alpha$ of the previous plots in (c), (f), (i).}
\label{infsize(r=0)}
\end{center}
\end{figure}

We first examine the fraction of infected nodes at the steady state
$\rho(\infty)$ as a function of $\beta$ with a small fraction of seeds
$\rho(0)=0.01$ and a large fraction of seeds
$\rho(0)=0.99$. Figures~\ref{infsize(r=0)}(a), \ref{infsize(r=0)}(d),
and \ref{infsize(r=0)}(g) show the results for three typical values
$\alpha=-2.0$, $-1.0$, and $1.0$, respectively. We find the following:

\begin{itemize}

\item[{(i)}] The value of $\rho(\infty)$ increases continuously with
  $\beta$ for the three values of $\alpha$ when $\rho(0)=0.01$ and
  $\rho(0)=0.99$.

\item[{(ii)}] When $\alpha=-2.0$ and $\alpha=-1.0$, there are two phase transitions
  \cite{colomer2014double,chen2014microtransition} of $\rho(\infty)$ for
  $\rho(0)=0.99$ and a single
  phase transition for $\rho(0)=0.01$ [see figures~\ref{infsize(r=0)} (a)
  and \ref{infsize(r=0)} (d)].
  When $\alpha=1.0$ there is a single phase transition for both
  $\rho(0)=0.01$ and $\rho(0)=0.99$ [see figure~\ref{infsize(r=0)} (g)]. Figures~\ref{infsize(r=0)}(b),
  \ref{infsize(r=0)}(e), and \ref{infsize(r=0)}(h) show peaks of $\chi$
  that are transition points for $\rho(0)=0.99$ (blue squares) and
  $\rho(0)=0.01$ (red line).

\item[{(iii)}] The plot indicates two hysteresis loops when
  $\alpha=-2.0$ and $\alpha=-1.0$, and a single hysteresis loop when
  $\alpha=1.0$.  Here we denote by $\beta_{\rm inv}$ the invasion
  threshold when $\rho(0)=0.01$, and $\beta_{\rm per}$ the persistence
  threshold when $\rho(0)=0.99$ \cite{gross2006epidemic}. In addition,
  we denote $\beta_{\rm per}^{I}$ and $\beta_{\rm per}^{II}$ the first
  and the second invasion (persistence) thresholds.

\end{itemize}

\begin{figure}[t]
\begin{center}
\includegraphics[width=1.0\linewidth]{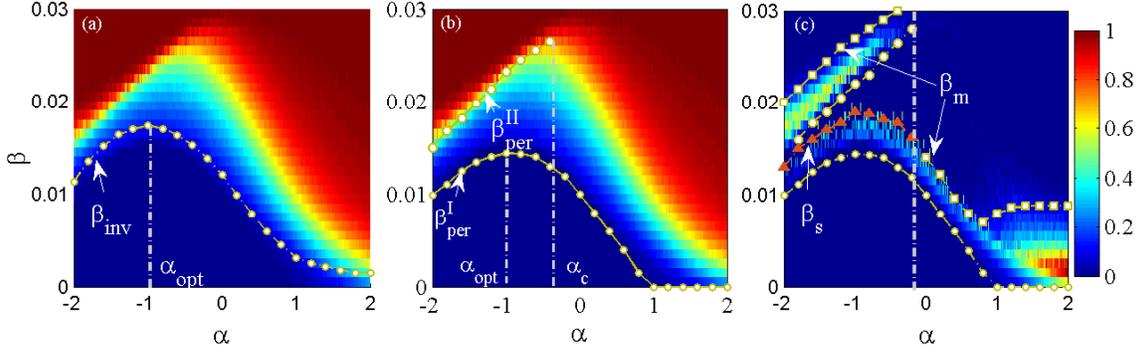}
\caption{(Color online) Dependence of $\rho(\infty)$ on $\beta$ and
  $\alpha$ when $r=0$. Color-coded values of epidemic size obtained from
  simulations for $\rho(0)=0.01$ (a) and $\rho(0)=0.99$ (b).  (c) The
  difference between the value of $\rho(\infty)$ in (a) and (b). The
  yellow circles are the numerical prediction of the invasion threshold
  $\beta_{inv}$ and the persistence threshold $\beta_{\rm per}$
  respectively. Red triangles and yellow squares represent the two
  bifurcation points $\beta_s$ and $\beta_m$ respectively.  The vertical
  dotted line in (a) indicates the location of the optimal value
  $\alpha_{opt}$, and in (b) and (c) indicates the location of critical
  value $\alpha_c$.  }
\label{phasediagram(r=0)}
\end{center}
\end{figure}

We next examine the underlying mechanism of the hysteresis
loop. Figures~\ref{infsize(r=0)}(c), \ref{infsize(r=0)}(f), and
\ref{infsize(r=0)}(i) show the ensemble average recovery rate at the
steady state $\langle\mu(\infty)\rangle=1/N\sum \mu_{i}(\infty)$, for
$\alpha=-2.0$, $\alpha=-1.0$, and $\alpha=1.0$, respectively.  We find
that for these values of $\alpha$, prior to the threshold the average
recovery rate is $\langle\mu(\infty)\rangle=1.0$ and after the threshold
it decreases continuously with $\beta$.  When the spreading process
begins with a low fraction of seeds, i.e., $\rho(0)=0.01$, the recovery
rate is higher than when there is a larger initial fraction of seeds,
i.e., $\rho(0)=0.99$ [see Figs.~\ref{infsize(r=0)}(c),
  \ref{infsize(r=0)}(f), and \ref{infsize(r=0)}(i)]. This is because
when $\rho(0)$ is small the fraction of susceptible nodes $(1-
\rho(0))$, is sufficiently high to generate a large number of
resources. A lower recovery rate for $\rho(0)=0.99$ delays the recovery
of infected nodes and increases the infection rate
$\lambda=\beta/\langle\mu(\infty)\rangle$
\cite{pastor2015epidemic}. Thus the disease breaks out at a lower
threshold when $\rho(0)=0.99$, and the value of $\rho(\infty)$ is larger
than when $\rho(0)=0.01$. Consequently there is a hysteresis loop.
In addition, when $\alpha=-2.0$ and $\alpha=-1.0$ the two curves of
$\langle\mu(\infty)\rangle$ for $\rho(0)=0.99$ and $\rho(0)=0.01$
overlap at some value of $\beta$ that separates the parameter space of
$\beta$ into two regions. Thus there are two hysteresis loops in the
separated regions.

\begin{figure}[t]
\begin{center}
\includegraphics[width=1.0\linewidth]{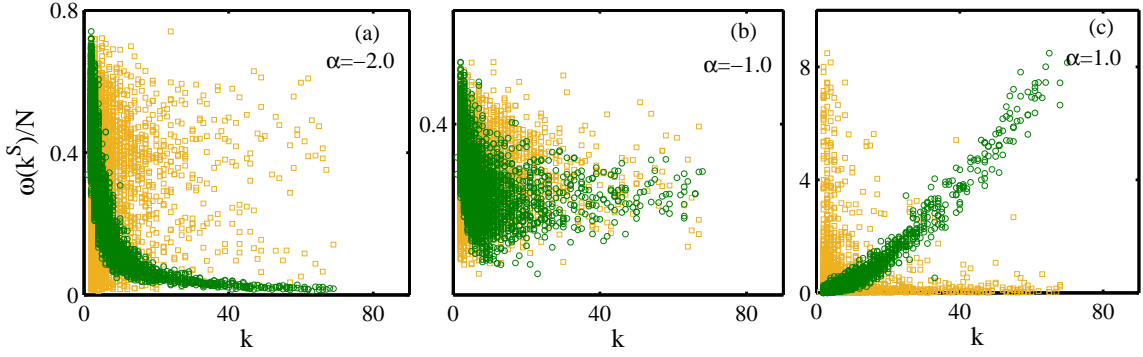}
\caption{(Color online) Scatter plots of resource quantity at
  $\beta=(\beta_{inv})_{-}$ for $\alpha=-2.0$ (a), $\alpha=-1.0$ (b) and
  $\alpha=1.0$ (c) when the inter-layer degree correlation $r=0$. The
  green circles represent scaled value of resource quantity
  $\omega(k^{\mathcal{S}})/N$ versus degree of nodes $k^{\mathcal{S}}$,
  and the yellow squares represent $\omega(k^{\mathcal{S}})/N$ versus
  the degree of the counterpart nodes $k^{\mathcal{C}}$.  The initial
  fraction of infected nodes is set to $\rho(0)=0.01$.}
\label{infsource(r=0)}
\end{center}
\end{figure}

To determine how preferential resource diffusion affects the dynamics of
disease spreading, we examine $\rho(\infty)$ as a function of $\beta$
and $\alpha\in[-2.0,2.0]$.  Figures~\ref{phasediagram(r=0)}(a) and
\ref{phasediagram(r=0)}(b) show the phase diagrams with initial
conditions $\rho(0)=0.01$ and $\rho(0)=0.99$, respectively, and
Fig.~\ref{phasediagram(r=0)}(c) shows the difference between values of
$\rho(\infty)$ in \ref{phasediagram(r=0)}(a) and
\ref{phasediagram(r=0)}(b). Note that $\rho(\infty)$ increases
continuously with $\beta$ at each fixed $\alpha$. In addition, when
$\rho(0)=0.01$ there is a single phase transition with one threshold
$\beta_{\rm inv}$ [circles in Fig.~\ref{phasediagram(r=0)}(a)]. When
$\rho(0)=0.99$ there is a critical $\alpha_c$ value below which there is
a double phase transition with two transition points $\beta_{\rm
  per}^{I}$ and $\beta_{\rm per}^{II}$ [circles in
  Fig.~\ref{phasediagram(r=0)}(b)]. Note that the thresholds in
Figs.~\ref{phasediagram(r=0)}(a) and \ref{phasediagram(r=0)}(b) are the
peaks of susceptibility $\chi$.  We also find that when $\beta$ is
fixed, $\rho(\infty)$ first decreases and then increases with $\alpha$
when $\beta$ is large, i.e., $\beta>\beta_{\rm inv}$ ($\beta>\beta_{\rm
  per}^{I}$) if there are two thresholds), and we obtain the minimum
value at the $\alpha_{\rm opt}$ where there is optimal resource
diffusion that optimally suppresses disease spreading. Note also that
the invasion threshold $\beta_{\rm inv}$ and persistence threshold
$\beta_{\rm per}^{I}$ [circles in (a) and (b)] have peak values at
$\alpha_{\rm opt}=-1.0$, which indicates an optimal resource diffusion
at $\alpha=-1.0$. Figure~\ref{phasediagram(r=0)}(c) shows that there are
two bifurcation points $\beta_s$ (triangles) and $\beta_m$ (squares),
and when $\alpha<\alpha_c$ there are two hysteresis loops in regions
[$\beta_{\rm per}^{I}, \beta_s$) and [$\beta_{\rm per}^{II}, \beta_m$). When
$\alpha>\alpha_c$ there is one hysteresis loop in region [$\beta_{\rm
  per}^{I}, \beta_m$).

To further explore these results, we study the resource distribution
(green circles) in layer $\mathcal{S}$ at the steady state when
$\beta=(\beta_{\rm inv})_{-}$ for $\rho(0)=0.01$, where
$\beta=(\beta_{\rm inv})_{-}$ is the infection rate immediately below
the threshold $\beta_{\rm inv}$ [see Fig.~\ref{infsource(r=0)}]. When
$\rho(0)=0.99$ we see similar results. Here we denote
$\omega(k^{\mathcal{S}},\infty)$ the resource quantity of nodes with
degree $k^{\mathcal{S}}$ at the steady state, where $k^{\mathcal{S}}$ is
the degree of nodes in layer $\mathcal{S}$, and
$\omega(k^{\mathcal{S}},\infty)/N$ the scaled value of
$\omega(k^{\mathcal{S}},\infty)$. Note that
$\omega(k^{\mathcal{S}},\infty)$ is shortened to
$\omega(k^{\mathcal{S}})$. In addition, to determine how the resource
distribution in layer $\mathcal{S}$ influences the recovery of nodes in
layer $\mathcal{C}$ at each value of parameter $\alpha$, we examine how
resources are distributed in nodes whose counterparts in layer
$\mathcal{C}$ have $k^{\mathcal{C}}$ degrees, where $k^{\mathcal{C}}$ is
the degree of nodes in layer $\mathcal{C}$.
This allows us to observe the the change trend of recovery rate with $\alpha$.

Figure~\ref{infsource(r=0)}(a) shows that when $\alpha=-2.0$ resources
move preferentially to low-degree nodes and $\omega(k^{\mathcal{S}})$ as
expected decays rapidly with $k^{\mathcal{S}}$. In addition, most of the
nodes in the two subnetworks with highly skewed degree distributions are
low-degree and only a few are high-degree. Thus the counterparts of the
high-degree nodes in layer $\mathcal{C}$ have a higher probability of
being low-degree nodes in layer $\mathcal{S}$ because of the random
correlation between the two layers. Thus most of the counterparts to the
high-degree nodes in layer $\mathcal{C}$ have large values of
$\omega(k^{\mathcal{S}})$ in layer $S$ [yellow squares in
  Fig.~\ref{infsource(r=0)}(a)], i.e., most high-degree nodes in layer
$\mathcal{C}$ have a high recovery rate that delays outbreaks of the
disease as $\beta$ increases. When $\alpha=1.0$ resources move
preferentially toward high-degree nodes in layer $\mathcal{S}$ and
agglomerate on high-degree nodes at the steady state. When there is a
random correlation between the two layers, most high-degree nodes
correspond to low-degree nodes in layer $\mathcal{C}$.  Thus the
resources of $k$-degree nodes in layer $\mathcal{S}$ increase with
$k^{\mathcal{S}}$ [see Fig.~\ref{infsource(r=0)}(c)]. In contrast, the
$\omega(k^{\mathcal{S}})$ decreases sharply with $k^{\mathcal{C}}$,
which indicates that the recovery rate of the high-degree nodes in layer
$\mathcal{C}$ rapidly declines when $\beta$ increases and resources
decrease. This in turn increases the effective infection rate
$\lambda=\beta/\langle\mu(\infty)\rangle$ in the
system. Figure~\ref{phasediagram(r=0)} shows that a severely skewed
distribution of resources lowers the epidemic threshold and a large
fraction of nodes when $\alpha$ is large, i.e., $\alpha=1.0$.

When $\alpha=-1.0$, the diffusion of resources in layer $\mathcal{S}$ is
less biased than when $\alpha=-2.0$ or $\alpha=1.0$. We analyze
Eq.~(\ref{expectRes}) and find that although low-degree nodes still have
a small advantage of acquiring resources, high-degree nodes can acquire
approximately the same quantity of resource at each time step because
they have more connections than low-degree nodes.  Thus resources are
distributed evenly for both high-degree and low-degree nodes [see
  Fig.~\ref{infsource(r=0)}(b)].  When resource diffusion is optimal,
all nodes in layer $\mathcal{C}$ have a rapid recovery rate [see
  Fig.~\ref{infsize(r=0)}(f)] that reduces the infection probability
between each pair of susceptible and infected nodes. Here the disease is
suppressed to the greatest extent. Figure~\ref{phasediagram(r=0)} shows
that the highest epidemic threshold $\beta_{\rm inv}$ ($\beta_{\rm
  per}$) and lowest fraction of infected nodes $\rho(\infty)$ are
obtained when resource diffusion is optimal, i.e., when $\alpha=-1.0$.

\section{Effect of inter-layer degree correlations on spreading dynamics}
\noindent
There are extensive interlayer correlations in real-world multiplex
systems \cite{kivela2014multilayer,LucasCorrelation}. In social
networks, for example, an individual with many daily face-to-face
contacts with colleagues tends to also have many social network contacts
\cite{szell2010multirelational}. In transportation networks, hub
airports tend to correlate with hub rapid transit stations
\cite{radicchi2015percolation}. We here investigate how the degree
correlations between the two layers impact the process of resource
diffusion and the dynamics of disease spreading. To construct a
double-layer correlated network with an adjustable degree of inter-layer
correlation, we first generate two subnetworks of the same size $N=5000$ and
the same power exponent $\gamma=2.2$ with a maximum positive or maximum
negative correlation. We then rematch each pair of counterpart nodes
with a probability $q$.  Thus the interlayer correlation after
rematching becomes \cite{lee2012correlated,wang2014asymmetrically}
\begin{equation}
r=\mid1-q\mid.
 \label{correlation}
\end{equation}
When the two layers are initially at maximum positive correlation
$r\geq0$, otherwise $r\leq0$.

\begin{figure}[H]
\begin{center}
\includegraphics[width=1.0\linewidth]{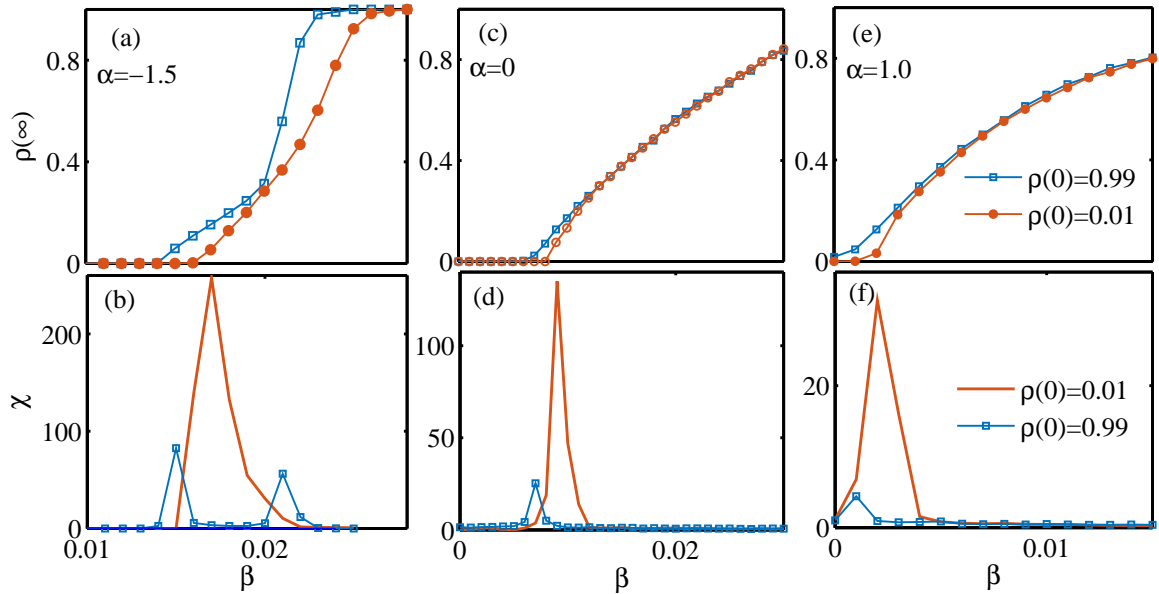}
\caption{(Color online) Influence of preferential resource diffusion on
  disease spreading when degree of inter-layer correlation is $r=-0.8$.
  $\rho(\infty)$ as a function of $\beta$ for $\alpha=-1.5$ (a),
  $\alpha=0$ (c) and $\alpha=1.0$ (e) respectively, Initial condition is
  set to $\rho(0)=0.01$ (red circles) and $\rho(0)=0.99$ (blue squares)
  respectively in the figures.  Susceptibility measure $\chi$ as a
  function of $\beta$ for $\alpha=-1.5$ (b), $\alpha=0$ (d) and
  $\alpha=1.0$ (f) respectively.}
\label{infsize(capa_-0.8)}
\end{center}
\end{figure}

\begin{figure}[H]
\begin{center}
\includegraphics[width=1.0\linewidth]{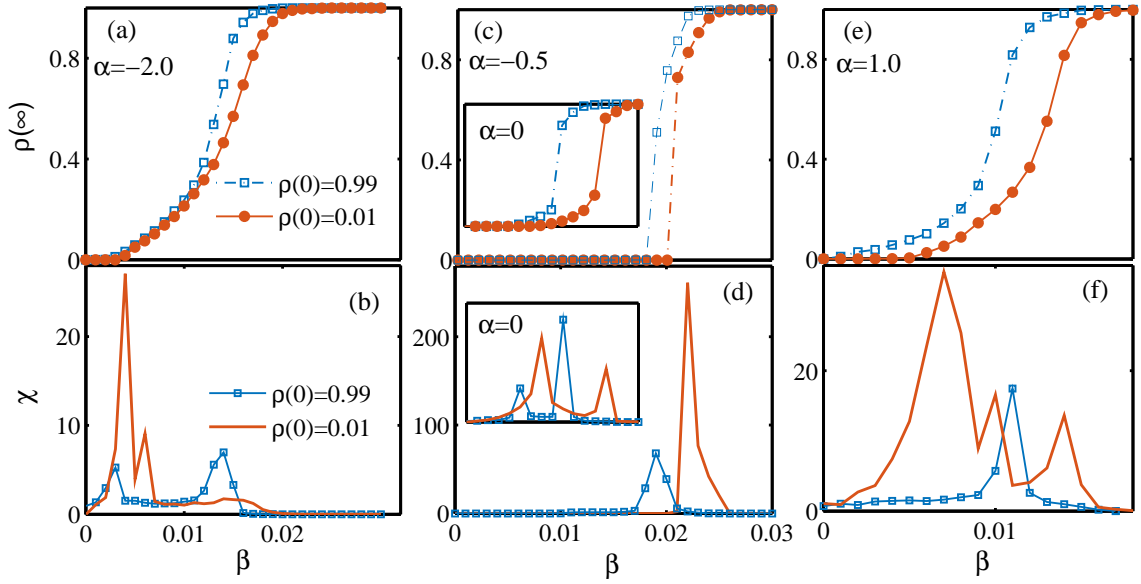}
\caption{(Color online) Influence of preferential resource diffusion on
  disease spreading when degree of inter-layer correlation $r=0.8$.
  $\rho(\infty)$ as a function of $\beta$ for $\alpha=-2.0$ (a),
  $\alpha=-1.0$ (c) and $\alpha=1.0$ (e) respectively.  Inset of (b) is
  $\rho(\infty)$ vs. $\beta$ for $\alpha=0$, Initial condition is set to
  $\rho(0)=0.01$ (red circles) and $\rho(0)=0.99$ (blue squares)
  respectively in the figures.  Susceptibility measure $\chi$ as a
  function of $\beta$ for $\alpha=-2.0$ (b), $\alpha=-1.0$ (d) and
  $\alpha=1.0$ (f) respectively. Inset of (d) is $\chi$ vs. $\beta$ for
  $\alpha=0$.  }
\label{infsize(capa_0.8)}
\end{center}
\end{figure}

Figure~\ref{infsize(capa_-0.8)} shows $\rho(\infty)$ as a function of
$\beta$ when there is a large negative interlayer correlation, i.e.,
$r=-0.8$. Figure~\ref{infsize(capa_0.8)} shows the same when
$r=0.8$. When $r=-0.8$ note the results of three typical values
$\alpha=-1.5$, $0$, and $1.0$ for $\rho(0)=0.01$ (red circles) and
$\rho(0)=0.99$ (blue squares). When $\alpha=-1.5$,
$\rho(\infty)$ has two phase transitions for $\rho(0)=0.99$ and two
hysteresis loops [see Figs.~\ref{infsize(capa_-0.8)}(a) and
  \ref{infsize(capa_-0.8)}(b)]. When $\alpha=0$ and $\alpha=1.0$,
$\rho(\infty)$ has one phase transition and a single hysteresis
loop. The peak values of $\chi$ in Figs.~\ref{infsize(capa_-0.8)}(b),
\ref{infsize(capa_-0.8)}(d), and \ref{infsize(capa_-0.8)}(f) are the
transition points for $\rho(0)=0.01$ (red lines) and $\rho(0)=0.99$
(blue squares).

Figure~\ref{infsize(capa_0.8)} shows the four typical values
$\alpha=-2.0$, $-0.5$, 0, and 1.0 when $r=0.8$. We find that when
$\alpha$ increases, the phase transition of $\rho(\infty)$ changes from
multiple continuous [$\alpha=-2.0$, see Fig.~\ref{infsize(capa_0.8)}(a)]
to discontinuous [Fig.~\ref{infsize(capa_0.8)}(c)] to hybrid [inset of
  \ref{infsize(capa_0.8)}(c)]. Eventually it returns to being multiple
continuous [Fig.~\ref{infsize(capa_0.8)}(e)]. In addition, when
$\alpha=1.0$ the first threshold disappears when $\rho(0)=0.99$. Later
we will use a finite-size scaling analysis to demonstrate the
discontinuous increase of $\rho(\infty)$
\cite{newman1999monte,radicchi2010explosive,chen2016crossover}.  Note
that, unlike when $r=0$ or $r=-0.8$, there is single hysteresis loop for
all values of $\alpha$. We can obtain the same explanation for the
hysteresis loops by analyzing the ensemble average recovery rate
$\langle\mu(\infty)\rangle$ as a function of $\beta$, similar to when
$r=0$.

\begin{figure}[H]
\begin{center}
\includegraphics[width=1.0\linewidth]{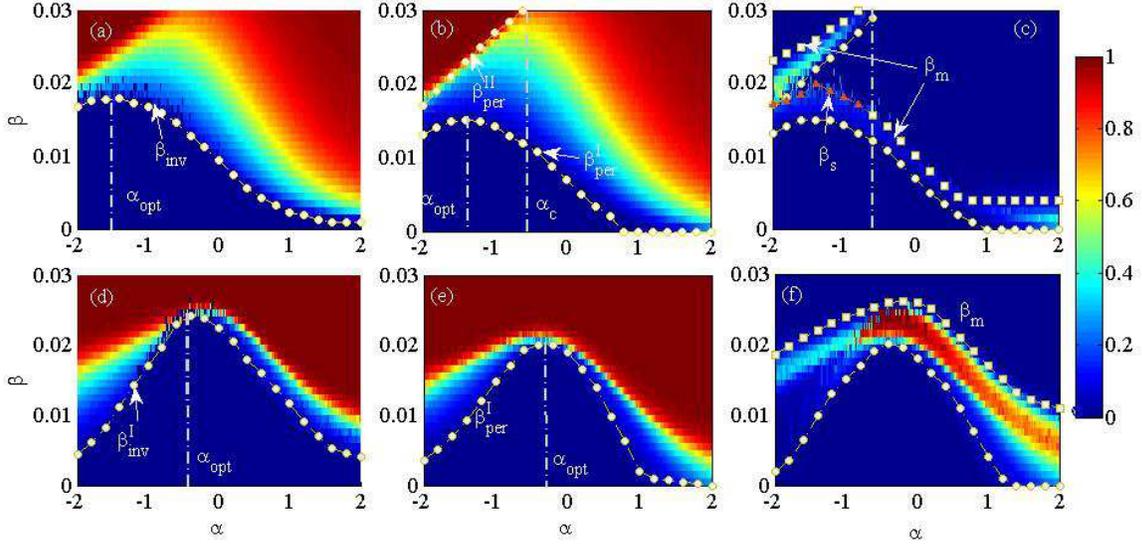}
\caption{(Color online) Dependence of $\rho(\infty)$ on $\beta$ and
  $\alpha$ when $r=-0.8$ (the first row) and $r=0.8$ (the second row).
  Color-coded values of epidemic size obtained from simulations for
  $\rho(0)=0.01$ (a), (d) and $\rho(0)=0.99$ (b), (e). The difference of
  the value of $\rho(\infty)$ in (a), (b) and (d), (e). The yellow
  circles are the numerical prediction of the invasion threshold
  $\beta_{inv}$ and the persistence threshold $\beta_{\rm per}$
  respectively, which are obtained from the peaks of the susceptibility
  measure $\chi$. Triangles and squares in (c), (f) represent the
  bifurcation points $\beta_s$ and $\beta_m$ respectively. The vertical
  dashed lines in (a), (d) indicate the location of the optimal value
  $\alpha_{opt}$, and in (b), (c) indicate the location of critical
  value $\alpha_c$. }
\label{alpha3D(combine)}
\end{center}
\end{figure}

To determine how preferential resource diffusion affects the dynamics of
epidemic spreading when there is interlayer degree correlation, we use
two-parameter $(\alpha,\beta)$ phase diagrams for $r=-0.8$ and $r=0.8$
[see Fig.~\ref{alpha3D(combine)}]. The
colors used in the figures are the values of $\rho(\infty)$. We set the
initial fraction of seeds at $\rho(0)=0.01$ in
Figs.~\ref{alpha3D(combine)}(a) and \ref{alpha3D(combine)}(d) and at
$\rho(0)=0.99$ in Figs.~\ref{alpha3D(combine)}(b) and
\ref{alpha3D(combine)}(e) at $r=-0.8$ and $r=0.8$,
respectively. Figures~\ref{alpha3D(combine)}(c) and
\ref{alpha3D(combine)}(f) show the differences between $\rho(\infty)$ in
Figs.~\ref{alpha3D(combine)}(a) and \ref{alpha3D(combine)}(b) and in
Figs.~\ref{alpha3D(combine)}(c) and \ref{alpha3D(combine)}(d).  Note
that there are optimal values of $\alpha$, i.e., $\alpha_{\rm
  opt}\simeq-1.5$ for $r=-0.8$ [see Figs.~\ref{alpha3D(combine)}(a) and
  \ref{alpha3D(combine)}(b)] and $\alpha_{\rm opt}\simeq-0.5$ for
$r=0.8$ [see Figs.~\ref{alpha3D(combine)}(d) and
  \ref{alpha3D(combine)}(e)].  Around $\alpha_{\rm opt}$ the disease is
maximally suppressed, the value of $\beta_{\rm inv}$ ($\beta_{\rm per}$)
reaches a maximum, and $\rho(\infty)$ a minimum [see
  Figs.~\ref{alpha3D(combine)}(a) and \ref{alpha3D(combine)}(b) and
  Figs.~\ref{alpha3D(combine)}(c) and \ref{alpha3D(combine)}(d) for
  $r=-0.8$ and $r=0.8$, respectively].  Similar to when $r=0$, when
$r=-0.8$ and $\rho(0)=0.99$ there is an $\alpha_c$ critical value. When
$\alpha<\alpha_c$ there are two phase transitions of $\rho(\infty)$ with
two transition points $\beta_{\rm per}^{I}$ and $\beta_{\rm per}^{II}$
[see Fig.~\ref{alpha3D(combine)}(b)]. When $\alpha>\alpha_c$ the
transition of $\rho(\infty)$ becomes single-phase.  When $\rho(0)=0.01$
there is a single phase transition of $\rho(\infty)$ [see
  Fig.~\ref{alpha3D(combine)}(a)].

We obtain thresholds from susceptibility
$\chi$. Figure~\ref{alpha3D(combine)}(c) shows that when
$\alpha<\alpha_c$ there are two bifurcations, $\beta_s$ (triangles) and
$\beta_m$ (squares) where $\beta_s<\beta_m$. There are two hysteresis
loops in regions [$\beta_{\rm per}^{I}, \beta_s$) and [$\beta_{\rm
  per}^{II}, \beta_m$). When $\alpha>\alpha_c$ there is one hysteresis
loop in region [$\beta_{\rm per}^{I}, \beta_m$). We find multiple phase
transitions when $r=0.8$ and when $\alpha$ is far from $\alpha_{\rm
  opt}$, i.e., $\alpha=-1.0$ or $\alpha=1.0$.  Note that for simplicity
we display only the first invasion threshold $\beta_{\rm inv}^{I}$ and
the first persistence threshold $\beta_{\rm per}^{I}$ in
Figs.~\ref{alpha3D(combine)}(d) and \ref{alpha3D(combine)}(e) (circles),
which we obtain from susceptibility measurement $\chi$. When $\alpha$
approaches $\alpha_{\rm opt}$, i.e., when $\alpha=-0.5$, the value of
$\rho(\infty)$ jumps from zero to a high value. In addition, the
difference in $\rho(\infty)$ values in Figs.~\ref{alpha3D(combine)}(d)
and \ref{alpha3D(combine)}(e) indicates the single hysteresis region
$(\beta_{\rm per}^{I}, \beta_m)$ [white circles and white squares in
  Fig.~\ref{alpha3D(combine)}(f)].

\begin{figure}[t]
\begin{center}
  \includegraphics[width=1.0\linewidth]{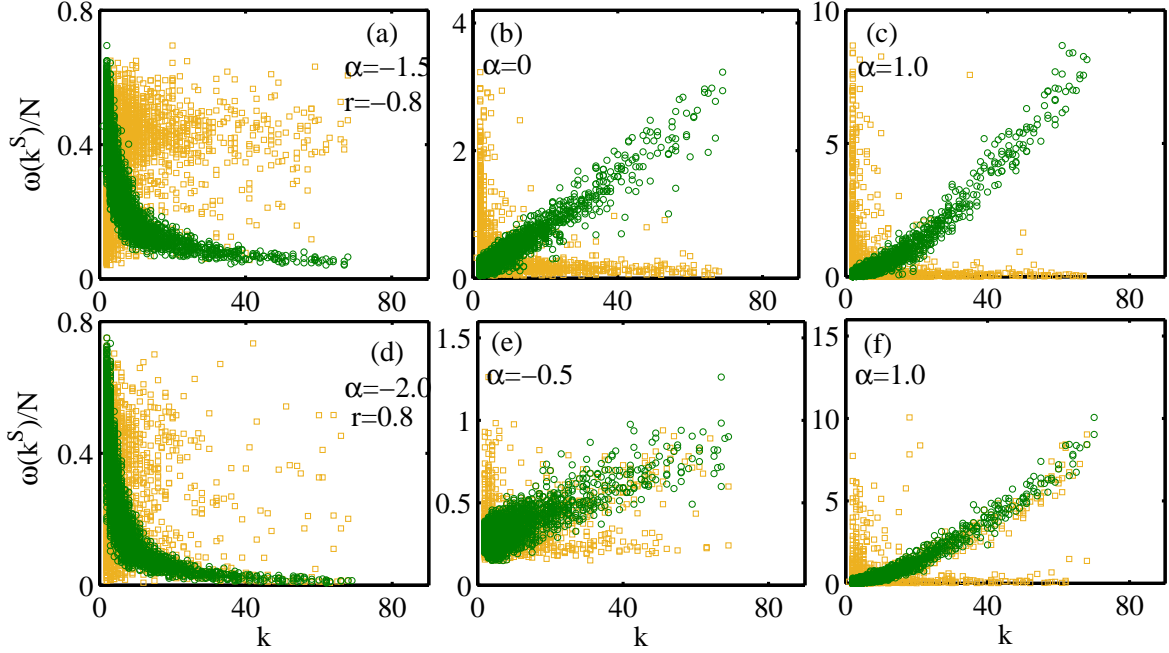}
\caption{(Color online) Scatter plots of resource quantity at
  $\beta=(\beta_{inv}^{I})_{-}$ when the inter-layer degree correlation
  $r=-0.8$ (a-c), and $r=0.8$ (d-f).  The green circles represent scaled
  resource quantity $\omega(k^{\mathcal{S}})/N$ versus
  $k^{\mathcal{S}}$, and the yellow squares represent
  $\omega(k^{\mathcal{S}})/N$ versus $k^{\mathcal{C}}$.  The initial
  fraction of infected nodes is set to $\rho(0)$ = 0.01.}
  \label{resource(combine)}
\end{center}
\end{figure}

To explain the optimization we examine the resource distribution of
nodes in layer $\mathcal{S}$ and how resources are distributed on those
nodes with counterparts in layer C that have $k^{\mathcal{C}}$ degrees
with initial $\rho(0)=0.01$ when $\beta=(\beta_{inv}^{I})_{-}$
[$\beta=(\beta_{inv})_{-}$ if it is a single phase transition].  Thus we
obtain the scatter plots of $\omega(k^{\mathcal{S}})/N$ versus
$k^{\mathcal{S}}$ (green circles) and $\omega(k^{\mathcal{S}})/N$ versus
$k^{\mathcal{C}}$ (yellow squares).  We obtain results similar to those
when $\rho(0)=0.99$.
Figures~\ref{resource(combine)}(a)--\ref{resource(combine)}(c) show
resource distributions for $\alpha=-1.5$, 0, and 1.0, respectively,
when $r=-0.8$. Note that when $\alpha=-1.5$ the probability that
resources move to low-degree nodes in layer $\mathcal{S}$ is
high. Figure~\ref{resource(combine)}(a) shows that
$\omega(k^{\mathcal{S}})/N$ decreases sharply when $k^{\mathcal{S}}$ in
layer $\mathcal{S}$ increases (green circles). In addition, when the
correlation between the two layers is negative, high-degree nodes in
layer $\mathcal{C}$ correlate with low-degree nodes. Because low-degree
nodes are more numerous in a heterogeneous network, most low-degree
nodes in layer $\mathcal{C}$ still have low-degree counterparts. Thus
both high-degree and low-degree nodes in layer $\mathcal{C}$ can rapidly
recover because there are adequate resources supplied by their
counterparts in layer $\mathcal{S}$ [yellow squares in
  Fig.~\ref{resource(combine)}(a)]. When this is the case, the disease
is effectively constrained [see Fig.~\ref{alpha3D(combine)}(a)].  When
$\alpha=0$ and $\alpha=1.0$, resources move preferentially to the few
high-degree nodes in layer $S$ and low-degree nodes receive little
[green circles in Figs.~\ref{resource(combine)}(b) and
  \ref{resource(combine)}(c)]. When $\beta$ increases, the recovery rate
of high-degree nodes in layer $\mathcal{C}$ rapidly decreases because
they cannot receive resources from their counterparts [yellow circles in
  Figs.~\ref{resource(combine)}(b) and \ref{resource(combine)}(c)] and
the disease is not constrained. Thus we see a small threshold and a
large $\rho(\infty)$ when resources move preferentially to high-degree
nodes in layer $\mathcal{S}$.

When $r=0.8$, to constrain disease spreading the recovery rate of both
high and low degree nodes in layer $\mathcal{C}$ should maintain a high
threshold.  To achieve this, resources must diffuse to high-degree nodes
in layer $\mathcal{S}$, i.e., $\alpha\simeq-0.5$, in a positive
correlation between the two layers. Thus when $\alpha=\alpha_{\rm
  opt}\simeq0.5$ there is a maximum threshold value and a minimum
$\rho(\infty)$ value when $\beta$ is fixed.

Figure~\ref{resource(combine)}(f) shows that when resources move only to
high-degree nodes in layer $\mathcal{S}$, i.e., when $\alpha=1.0$, there
are no resources for the low-degree nodes in layer
$\mathcal{S}$. Figure~\ref{resource(combine)}(d) shows that when
resources move only to low-degree nodes, there are none for the
high-degree nodes. In both of these extreme conditions, the node
recovery rate in layer $\mathcal{C}$ declines rapidly as $\beta$
increases, which causes an earlier outbreak of disease [see
  Figs.~\ref{alpha3D(combine)}(d) and \ref{alpha3D(combine)}(e)].

\begin{figure}[t]
\begin{center}
\includegraphics[width=1.0\linewidth]{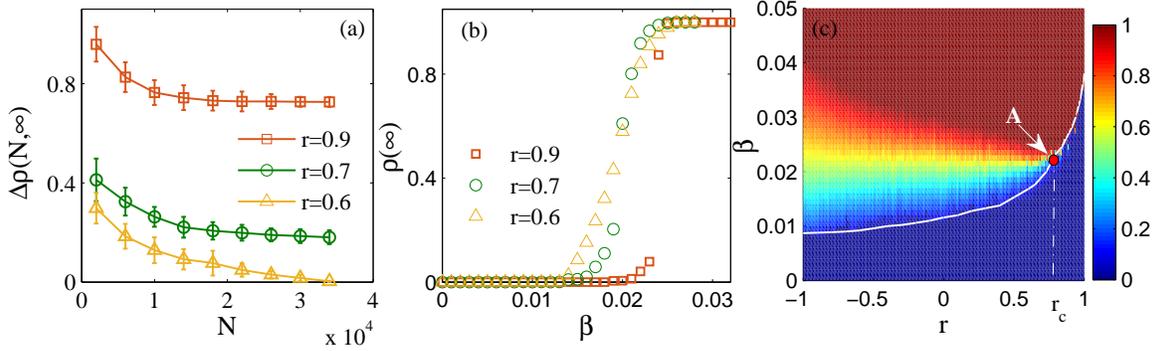}
\caption{(Color online) Results of finite-size scaling analysis for the
  discontinuous increase of $\rho(\infty)$. (a) Increase of infected
  density at the steady state $\Delta\rho(N,\infty)$ as a function of
  network size $N$ for $r=0.9$ (red squares), $r=0.7$ (green circles)
  and $r=0.6$ (yellow triangles). (b) Infected density $\rho(\infty)$
  as a function of $\beta$ for $r=0.9$ (red squares), $r=0.7$ (green
  circles) and $r=0.6$ (yellow triangles) respectively. (c)
  Dependence of $\rho(\infty)$ on $r$ and $\beta$. Color-coded values
  of $\rho(\infty)$ obtained from simulations with initial condition
  $\rho(0)=0.01$. Point $A$ is a triple point and the corresponding
  $r$ is the critical value $r_c$. White line represents the first
  epidemic threshold $\beta_{inv}^{I}$ that are obtained from the
  peaks of $\chi$. The bias parameter is set to $\alpha=0$.}
\label{finite}
\end{center}
\end{figure}

We next use a finite-size scaling analysis to examine the discontinuous
increase of $\rho(\infty)$ when $\alpha$ approaches $\alpha_{\rm opt}$
and the two network layers are positively correlated.  We define
$\rho(N,\infty)$ the fraction of infected nodes at the steady state for
a network with $N$ nodes and $\Delta\rho(N,\infty)$ the maximum increase
of $\rho(N,\infty)$ during an infinitely small increase of $\beta$,
which is expressed
\begin{equation}
  \Delta\rho(N,\infty)=max_{\beta\in[0,1]}
\{\rho(N,\infty,\beta+\Delta\beta)-\rho(N,\infty,\beta)\},
  \label{jumpRho}
\end{equation}
where $\Delta\beta$ is an infinitesimal increment of $\beta$, set at
$\Delta\beta=0.001$ in our simulations, and $\rho(N,\infty,\beta)$ is
the fraction of infected nodes at steady state when infection rate is
$\beta$. When
\begin{equation}
 \lim_{N\rightarrow\infty}\Delta\rho(N,\infty)>0.0,
  \label{limit}
\end{equation}
there is a discontinuous increase in $\rho(\infty)$
\cite{nagler2011impact,radicchi2015percolation}. Note that we use
$\alpha=0$ for the finite-size scaling analysis. Figure~\ref{finite}(a)
shows $\Delta\rho(N,\infty)$ as a function of $N$ when $\alpha=0.6$
(orange triangles), $\alpha=0.7$ (green circles), and $\alpha=0.9$ (red
squares). Note that when $\alpha=0.6$, $\Delta\rho(N,\infty)$ converges
to $0$ asymptotically. When $\alpha=0.7$ and $\alpha=0.9$,
$\Delta\rho(N,\infty)$ asymptotically converges to a positive constant.

Figure~\ref{finite}(b) shows $\rho(\infty)$ as a function of $\beta$
when $\rho(0)=0.01$ for three typical values of interlayer correlation
$r = 0.6$, $r=0.7$, and $r=0.9$ in a network of
size $N=10000$. Note that when $r=0.6$, $\rho(\infty)$ increases
continuously with $\beta$. When $r=0.7$ and $r=0.9$, $\rho(\infty)$
first increases slowly and continuously at $\beta_{\rm inv}^{I}$, and
then jumps discontinuously at $\beta_{\rm inv}^{II}$, all of which are
characteristics of a hybrid phase transition.

We next use extensive simulations to obtain the phase diagram of
$\rho(\infty)$ in the two-parameter $(r,\beta)$ plane with an initial
condition $\rho(0)=0.01$ when $\alpha=0$. When $\rho(0)=0.99$ the
results are similar. Figure~\ref{finite} (c) shows that when the
two layers are negatively correlated ($r<0$), $\rho(\infty)$ increases
continuously with $\beta$. When $r>0$, there is a critical value
point $r_c$ [point $A$ in Fig.~\ref{finite} (c)]. When $r\geq r_c$
there is a discontinuous change of $\rho(\infty)$ at the threshold.
Note also that the epidemic threshold increases with $r$ [see white line
in Fig.~\ref{finite} (c)].

\begin{figure}[H]
\begin{center}
\includegraphics[width=0.5\linewidth]{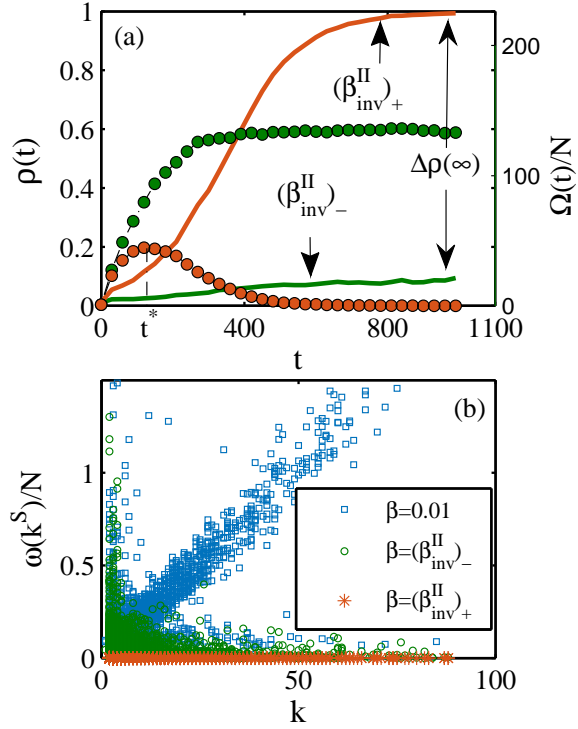}
\caption{(Color online). Analysis of the hybrid phase transition.  (a)
  The left vertical axis shows the time evolution of $\rho(t)$ when
  $\beta$ is just below the second threshold $(\beta_{\rm
    inv}^{II})_{-}$ (the lower green line) and just over $(\beta_{\rm
    inv}^{II})_{+}$ (the upper red line). The right vertical axis shows
  the time evolution of scaled total resources of all nodes
  $\Omega(t)/N$ for $(\beta_{\rm inv}^{II})_{-}$ (the upper green
  circles) and $(\beta_{\rm inv}^{II})_{+}$ (the lower red circles). (b)
  Resource distribution in layer $S$ for $\beta=0.01$ (blue squares),
  $(\beta_{\rm inv}^{II})_{-}$ (green circles) and $(\beta_{\rm
    inv}^{II})_{+}$ (red stars).}
\label{timeevol}
\end{center}
\end{figure}

To explain the hybrid discontinuous phase transition, we plot the time
evolution of total resources $\Omega(t)$ and infected fraction $\rho(t)$
with the initial condition $\rho(0)=0.01$ for $r=0.9$ when $\alpha=0$
[see Fig.~\ref{timeevol}(a)]. When $\rho(0)=0.99$ the results are
similar. Figure~\ref{timeevol}(b) shows the corresponding resource
distribution at the steady state. When $\beta$ is immediately below
$(\beta_{\rm inv}^{II})_{-}$, the scaled value of the total resources
$\Omega(t)/N$ abruptly increases at the early stage of the diffusion
process [green circles in Fig.~\ref{timeevol}(a)] because almost all
nodes in layer $\mathcal{C}$ are healthy and resources are constantly
generated by the corresponding nodes in layer $\mathcal{S}$.  After a
longer period of time $t>300$ the system enters a steady state, and
fluctuations stay within a small range (upper green circles). Here the
resources of high-degree nodes are rapidly consumed, and the resource
level for low-degree nodes remains high [see Fig.~\ref{timeevol}(b)]
indicating that the disease is localized around the high-degree
nodes. We thus learn that before $\beta_{\rm
  inv}^{II}$ the system changes from a disease-free
absorbing phase to a locally active phase (in which $\rho(\infty)$
reaches a finite small value) at $\beta_{\rm inv}^{I}$ [green line in
  Fig.~\ref{timeevol}(a)]. For the sake of comparison,
Fig.~\ref{timeevol}(b) shows a plot of the resource distribution when
$\beta=0.01$.

When $\beta=(\beta_{\rm inv}^{II})_{+}$, the value of $\Omega(t)/N$
rapidly increases as the disease spreads from the local area of the
seeds [red circles in Fig.~\ref{timeevol}(a)]. As $t$ increases
$\rho(t)$ slowly increases and $\Omega(t)/N$ reaches a peak value at a
crossover time $t^*$. After $t^*$, $\Omega(t)/N$ drops rapidly,
indicating that the newly-generated node resources in layer
$\mathcal{S}$ are not sufficient to recover the infected nodes in layer
$\mathcal{C}$. The recovery rate of the infected nodes then declines as
resources decrease, which induces an increase in the infection rate of
the disease, especially in the hub nodes. Thus as the infection rate
increases, the resources available in layer $\mathcal{S}$ further
decrease and the node recovery rate in layer $\mathcal{C}$ decreases.
Then a cascading effect appears that sharply increases $\rho(t)$ from a
small finite value to a value near $1.0$ [red line in
  Fig.~\ref{timeevol}(a)].  Figure~\ref{timeevol}(a) shows
$\Delta\rho(\infty)$, which is the increase of $\rho(\infty)$ when
$\beta$ increases from $(\beta_{\rm inv}^{II})_{-}$ to $(\beta_{\rm
  inv}^{II})_{+}$. This indicates a discontinuous jump in
$\rho(\infty)$.  Figure~\ref{timeevol}(b) shows that all resources of
all nodes in the network have been consumed, in contrast to when
$(\beta_{\rm inv}^{II})_{-}$.

\begin{figure}[H]
\begin{center}
\includegraphics[width=0.8\linewidth]{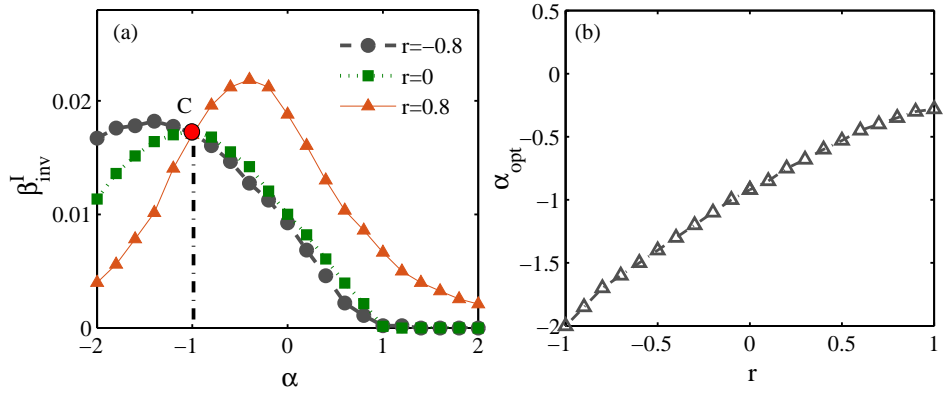}
\caption{(Color online) The value of the first invasion threshold
  $\beta_{\rm inv}^{I}$ as a function of the bias parameter $\alpha$ for
  $r=-0.8$ (gray circles), $r=0$ (green squares) and $r=0.8$ (red
  triangles) (a). Optimal bias parameter $\alpha_{\rm opt}$ as a
  function of inter-layer correlation $r$ (b).  Each symbol in (a) is
  obtained from the susceptibility measure. Red circle at $\alpha=-1.0$
  is the cross point of the three lines. Here the initial condition is
  set to $\rho(0)=0.01$.}
\label{threshold_opt}
\end{center}
\end{figure}

Figure~\ref{threshold_opt}(a) plots the value of the first invasion
threshold $\beta_{\rm inv}^{I}$ as a function of $\alpha$ for three
typical degree correlations, $r=-0.8$ (gray circles), $r=0$ (green
squares), and $r=0.8$ (red triangles), with an initial condition
$\rho(0)=0.01$. When $\rho(0)=0.99$ the results are similar.  Note that
the three curves cross at $\alpha=-1.0$ [point $C$ in
  Fig.~\ref{threshold_opt}(a)]. When $\alpha<-1$, $\beta_{\rm inv}^{I}$
decreases with $r$, but when $\alpha>-1$, $\beta_{\rm inv}^{I}$
increases with $r$. When $\alpha<-1$, resources move preferentially to
low-degree nodes in layer $S$. To suppress the spreading, the nodes in
layer $S$ must supply enough resources to high-degree nodes in layer
$C$. Thus negative interlayer correlation enhances the disease
suppression.  In contrast, when $\alpha>-1$ high-degree nodes add
resources in layer $S$. To constrain these high-degree nodes we must
have high-degree counterparts in layer $C$. Thus we increase $\beta_{\rm
  inv}^{I}$ with $r$.

Finally we explore the relationship among the optimal values of the bias
parameter $\alpha_{\rm opt}$ at which the disease is maximally
controlled. Figure~\ref{threshold_opt}(b) shows $\alpha_{\rm opt}$ as a
function of $r$. Note that the value of $\alpha_{\rm opt}$ increases
monotonically with $r$ because, with the increase of the interlayer
correlation, the probability that the large degree nodes in layer
$\mathcal{S}$ have counterparts with large degrees also increases. To
protect the large degree nodes in layer $\mathcal{C}$, resources in
layer $\mathcal{S}$ must diffuse preferentially to large degree
nodes. Thus $\alpha_{\rm opt}$ increases with $r$.

\section{Conclusions and discussions}
We have explored how preferential resource diffusion affects the
dynamics of disease spreading in correlated multiplex networks. We
assume that resources diffuse in the social contact layer and that the
disease is transmitted in the physical contact layer of the network. The
two dynamical processes are coupled such that the generation and
diffusion of resources in layer $\mathcal{S}$ are dependent on the state
of nodes in layer $\mathcal{C}$, and that the recovery of infected nodes
in layer $\mathcal{C}$ are dependent on the resources of their
counterparts in layer $\mathcal{S}$. To model the disease spreading in
layer $\mathcal{C}$, we propose a resource-based
susceptible-infected-susceptible (rSIS) model. Using extensive
simulations we find that preferential resource diffusion can change the
phase transition in $\rho(\infty)$, i.e, when the degree of interlayer
correlation $r$ is below a critical value, the transition $\rho(\infty)$
in $\rho(0)=0.99$ changes from two continuous phase transitions to one
single phase transition as the controlling parameter $\alpha$ increases.
Note that when $\rho(0)=0.01$ the transition of $\rho(\infty)$ is single
and continuous throughout the parameter space of $\alpha$. In addition,
there are hysteresis loops in the continuous phase transitions. There
are two hysteresis loops accompanied by two phase transitions and one
single hysteresis loop accompanied by one single phase transition of
$\rho(\infty)$. When $r$ is above the critical value, the phase
transition of $\rho(\infty)$ changes from multiple ($\alpha$ is too
large or too small) to discontinuous, and then becomes hybrid and
exhibits the properties of both continuous and discontinuous transitions
($\alpha$ is near the optimal value). Note that there is an optimal
resource diffusion at each fixed value of $r$. When the diffusion of
resources is optimal the threshold reaches a maximum and the disease can
be maximally suppressed.

In recent years constraining disease epidemics in human populations has
become a hot research topic and has attracted many workers across a
variety of fields. Most research has focused on ways of optimally
allocating limited public resources, but there has been little
examination of how the resource diffusion among the individuals affects
spreading dynamics. Our model fills this gap. There remain limits in our
model. For example, because the model is complex we have not yet
developed theoretical solutions, and thus theoretically obtaining an
optimal solution $\alpha_{\rm opt}$ would be an interesting and
important path for future research.

\section*{Acknowledgments}
This work was supported by the National Natural Science
Foundation of China under Grant Nos. 61673086 and 11575041,
the Fundamental Research Funds for the Central Universities under Grant
No. ZYGX2015J153. LAB
is supported by UNMdP and Agencia, Pict 0429/2013.
The Boston University Center for Polymer Studies is
supported by NSFGrants PHY--1505000, CMMI-1125290 and CHE--1213217, by DTRA GrantHDTRA1--14--1--
0017 and by DOE Contract DE-AC07-05Id14517.

\section*{References}
\bibliographystyle{iopart-num}
\bibliography{bibfile}
\end{document}